\documentclass{article}%
\usepackage{amsfonts}
\usepackage{amsmath}
\usepackage{cite}
\usepackage{amssymb}
\usepackage{charter}
\usepackage{graphicx}%
\providecommand{\U}[1]{\protect \rule{.1in}{.1in}}
\begin{document}

\title{High efficiency Four-Wave Mixing in a five-level atomic System based on the
two electromagnetically induced \mbox{transparency}}
\author{Zhi-ping Wang$\thanks{{\small Corresponding author. E-mail addresses:
wzping@mail.ustc.edu.cn, wzping\_0098@163.com}},$\ Shuang-xi Zhang$%
\thanks{{\small Corresponding author. E-mail addresses:
shuangxi@mail.ustc.edu.cn}}$\\{\small Department of Material Science
and Engineering, University of Science and Technology of China,
Hefei,Anhui 230026, China}} \maketitle

\begin{abstract}
We have analyzed a five-level $\wedge$-configuration Four-Wave Mixing (FWM)
scheme for obtaining a high-efficiency FWM based on the two
electromagnetically induced transparency. We find that the maximum FWM
efficiency is nearly 30\%, which is orders of magnitude larger than previous
schemes based on the two electromagnetically induced transparency. Our scheme
may provide a new possibility for technological applications such as nonlinear
spectroscopy at very low light intensity, quantum single-photon nonlinear
optics and quantum information science.



\end{abstract}

\section{Introduction}

As we know, the light will be seriously absorbed when it passes through the
optical medium, which is very bad for the conversion efficiency of the
nonlinear optical processes. However, the situation has been changed since
Harris and his co-workers discovered the novel phenomenon of the
electromagnetically induced transparency (EIT) in a three-level atomic system
in 1990s [1]. Then the approaches of using the EIT to control the absorptive
and dispersive properties of atomic medium attract the attentions of many
people. Meanwhile, as one of the centerpieces of modern technology, the
four-wave mixing (FWM) process based on the EIT in the ultraslow propagation
regime [2-11] has also attracted the attentions of many researchers for its
potential applications in nonlinear spectroscopy at very low light intensity,
quantum single-photon nonlinear optics, high-efficiency generation of
short-wavelength coherent radiation at pump intensity approaching the
single-photon level, and quantum information science. For instance, Harris et
al. proposed the use of EIT to suppress absorption of the short-wavelength
light generated in a FWM scheme and showed that the FWM efficiency could be
greatly enhanced [12]. Deng et al. proposed a FWM scheme based on EIT and
associated slow light propagation, and their calculations showed that a many
orders of magnitude increase in the FWM efficiency may be obtained [13]. Later
on, Wu et al. investigated and discussed a FWM scheme in a five-level atomic
system by the use of EIT, which led to suppressing both two-photon and
three-photon absorptions in FWM scheme and enabling the FWM to proceed through
real, resonant intermediate states without absorption loss [14], and then he
and his co-workers again analyzed a lifetime-broadened four-state FWM scheme
in the ultraslow propagation regime and put forward a new type of induced
transparency resulted from multiphoton destructive interference [15].

On the other hand, some persons began to study the FWM schemes based on the
two-EIT in the ultraslow propagation regime recently [6,16-18]. For example,
Gong et al. showed a tripod-configuration FWM scheme for enhancement of FWM
efficiency, and they found that the conversion efficiency of FWM enhanced
several orders by adjusting the intensity of the coupling fields [16]. Quite
recently, Huang et al. proposed a scheme to obtain a highly efficient FWM in a
coherent five-level tripod system by using a double-dark resonance and
multiphoton destructive interference induced transparency [17]. Otherwise,
Zhang and Xiao reported an experimental observation of optical
pumping-assisted FWM and two-EIT assisted SWM. They found that the efficient
SWM could be selected by EIT window and controlled by the coupling as well as
dressed field detuning and power. Due to the two-EIT and optical pumping
assistance, the enhanced SWM signal was more than ten times lager than the
coexisting FWM signal [18].

In this paper, we investigate a five-level $\wedge$-configuration system based
on the two electromagnetically induced transparency. Our study and the system
based on the Refs. [14-15,16-18], however, which are drastically different
from those works. First and foremost is that we are interested in showing the
effect of the external coherent driving field on both the relative intensity
of the generated FWM field and the conversion efficiency of FWM. Second, our
atomic system is a kind of atomic configuration, which owns the property of
the double-dark resonances. Third, an important advantage of our scheme is
that the high efficiency Four-Wave Mixing can be easily controlled by
adjusting the external coherent driving field. Our paper is organized as
follows: in Section 2, we present the theoretical model and establish the
corresponding Schr\"{o}dinger-Maxwell equations. Our numerical results and
physical analyses are shown in Section 3. In Section 4, some simple
conclusions are given.

\section{The model and the dynamic equations}

We consider the five-level $\wedge$-configuration system interacting with one
weak, pulsed pump field, two cw laser pump fields, and an external coherent
driving field as shown in Fig. 1. The transitions $\left \vert 3\right \rangle
\longrightarrow \left \vert 1\right \rangle $\ and $\left \vert 1\right \rangle
\longrightarrow \left \vert 2\right \rangle $\ are mediated by two laser pump
fields $\omega_{1}$ (Rabi frequency $2\Omega_{1}$) and $\omega_{2}$ (Rabi
frequency $2\Omega_{2}$) respectively. A weak, pulsed pump field $\omega_{p}$
(Rabi frequency $2\Omega_{p}$) and an external coherent driving field are
applied to the transitions $\left \vert 0\right \rangle \longrightarrow
\left \vert 3\right \rangle $\ and $\left \vert 3\right \rangle \longrightarrow
\left \vert 4\right \rangle $\ respectively.

In the interaction picture, with the rotating-wave approximation, the
interaction Hamiltonian of the system can be written as (we let $\hbar=1$)
[14,19,20],%
\[%
\begin{array}
[c]{l}%
H_{int}=\left(  \Delta_{p}-\Delta_{1}\right)  \left \vert 1\right \rangle
\left \langle 1\right \vert +\left(  \Delta_{p}-\Delta_{1}+\Delta_{2}\right) \\
\left \vert 2\right \rangle \left \langle 2\right \vert +\Delta_{p}\left \vert
3\right \rangle \left \langle 3\right \vert +\left(  \Delta_{p}-\Delta
_{c}\right)  \left \vert 4\right \rangle \left \langle 4\right \vert -(\Omega
_{c}e^{i\vec{k}_{c}\cdot \vec{r}}\\
\left \vert 3\right \rangle \left \langle 4\right \vert +\Omega_{1}e^{i\vec{k}%
_{1}\cdot \vec{r}}\left \vert 3\right \rangle \left \langle 1\right \vert
+\Omega_{2}e^{i\vec{k}_{2}\cdot \vec{r}}\left \vert 2\right \rangle \left \langle
1\right \vert +\\
\Omega_{p}e^{i\vec{k}_{p}\cdot \vec{r}}\left \vert 3\right \rangle \left \langle
0\right \vert +\Omega_{m}e^{i\vec{k}_{m}\cdot \vec{r}}\left \vert 2\right \rangle
\left \langle 0\right \vert +H.c.),\text{ \  \  \ (1)}%
\end{array}
\]

where $\Delta_{n}$ ($n=1,2,p,c$) represent the respective single-photon
detunings. $\overrightarrow{k}_{n}$ ($n=1,2,p,c,m$) is the respective wave
vector, and we assume $\Omega_{n}=\Omega_{n}^{\ast}$ ($n=1,2,p,c,m$).

From the Schr\"{o}dinger equation in the interaction picture, $i\partial
\left \vert \Psi \right \rangle /\partial t=H_{int}\left \vert \Psi \right \rangle
$, and defining the atomic state as,%

\begin{align}
\left \vert \Psi \right \rangle  &  =A_{0}\left \vert 0\right \rangle
+A_{1}e^{i\left(  \vec{k}_{p}-\vec{k}_{1}\right)  \cdot \vec{r}}\left \vert
1\right \rangle +\nonumber \\
&  A_{2}e^{i\left(  \vec{k}_{p}-\vec{k}_{1}+\vec{k}_{2}\right)  \cdot \vec{r}%
}\left \vert 2\right \rangle +A_{3}e^{i\vec{k}_{p}\cdot \vec{r}}\left \vert
3\right \rangle \nonumber \\
&  +A_{4}e^{i\left(  \vec{k}_{p}-\vec{k}_{c}\right)  \cdot \vec{r}}\left \vert
4\right \rangle , \label{2}%
\end{align}

Under rotating-wave and slowly varying envelope approximations, the dynamics
of atomic response and the optical field is governed by the
Maxwell-Schr\"{o}dinger equations,%

\[%
\begin{array}
[c]{l}%
-i\dot{A}_{1}=\Delta_{1}A_{1}+i\gamma_{1}A_{1}+\Omega_{1}^{\ast}A_{3}%
+\Omega_{2}^{\ast}A_{2}\\
-i\dot{A}_{2}=\Delta_{2}A_{2}+i\gamma_{2}A_{2}+\Omega_{2}A_{1}+\Omega_{m}%
A_{0}\\
-i\dot{A}_{3}=\Delta_{p}A_{3}+i\gamma_{3}A_{3}+\Omega_{p}A_{0}+\Omega_{1}%
A_{1}+\Omega_{c}A_{4}\\
-i\dot{A}_{4}=\Delta_{c}A_{4}+i\gamma_{4}A_{4}+\Omega_{c}^{\ast}A_{3}\text{
\  \  \  \  \  \  \  \  \  \ }\\
\frac{\partial \Omega_{p\left(  m\right)  }}{\partial z}+\frac{1}{c}%
\frac{\partial \Omega_{p\left(  m\right)  }}{\partial t}=i\kappa_{03\left(
02\right)  }A_{3\left(  2\right)  }A_{0}^{\ast}\text{ \  \  \  \  \ (3)}%
\end{array}
\]

in the above, $\gamma_{n}$ ($n=1,2,3,4$) are the decay rates. The
$\kappa_{03\left(  02\right)  }=2N\omega_{p\left(  m\right)  }\left \vert
D_{_{03\left(  02\right)  }}\right \vert ^{2}/\left(  c\hbar \right)  $ with
$N$\ and $D_{_{03\left(  02\right)  }}$\ being atomic density and dipole
moment between states $\left \vert 0\right \rangle $\ and $\left \vert
3\right \rangle $ ($\left \vert 2\right \rangle $), respectively, and we have
assumed the phase matching condition $\overrightarrow{k}_{n}$
$=\overrightarrow{k}_{p}+\overrightarrow{k}_{1}+\overrightarrow{k}_{2}$.

Following the method described in the Refs. [15,20], we take the Fourier
transform of Eq. (2) and the wave equations for the pulsed probe field
$\Omega_{p}$\ and FWM the generated field $\Omega_{m}$, and using the
undepleted ground-state approximation $A_{0}\approx1$, we can obtain%
\[%
\begin{array}
[c]{l}%
\left(  \omega+\Delta_{1}+i\gamma_{1}\right)  \alpha_{1}+\Omega_{2}^{\ast
}\alpha_{2}+\Omega_{1}^{\ast}\alpha_{3}=0\\
\left(  \omega+\Delta_{2}+i\gamma_{2}\right)  \alpha_{2}+\Omega_{2}\alpha
_{1}+\Lambda_{m}=0\\
\left(  \omega+\Delta_{p}+i\gamma_{3}\right)  \alpha_{3}+\Omega_{1}\alpha
_{1}+\Omega_{c}\alpha_{4}+\Lambda_{p}=0\\
\left(  \omega+\Delta_{c}+i\gamma_{4}\right)  \alpha_{4}+\Omega_{c}^{\ast
}\alpha_{3}=0\text{ \  \  \  \  \  \  \  \ }\\
\frac{\partial \Lambda_{p\left(  m\right)  }}{\partial z}-i\frac{\omega}%
{c}\frac{\partial \Lambda_{p\left(  m\right)  }}{\partial t}=i\kappa_{03\left(
02\right)  }\alpha_{3\left(  2\right)  }\text{ \  \  \  \  \  \ (4)}%
\end{array}
\]

here, $\Lambda_{p}$,\ $\Lambda_{m}$ and $\alpha_{j}$ ($j=1,2,3,4$) are the
Fourier transforms of $\Omega_{p}$,\ $\Omega_{m}$ and $A_{j}$ $(j=1,2,3,4),$ respectively.

Using the initial conditions $\Lambda_{p}\left(  0,\omega \right)  \neq
0$,\ $\Lambda_{m}\left(  0,\omega \right)  =0$ and solving the Eq. (3), it is
easily to obtain the following relations,%

\[%
\begin{array}
[c]{l}%
\alpha_{2}=\frac{-\Omega_{1}^{\ast}\Omega_{2}F_{4}}{D}\Lambda_{p}+\frac{D_{m}%
}{D}\Lambda_{m}\\
\alpha_{3}=\frac{-\Omega_{1}\Omega_{2}^{\ast}F_{4}}{D}\Lambda_{m}+\frac{D_{p}%
}{D}\Lambda_{p}\text{ \  \  \  \  \  \  \  \  \  \ }\\
\Lambda_{m}\left(  z,\omega \right)  =\Lambda_{p}\left(  0,\omega \right)
\frac{\kappa_{02}\Omega_{1}^{\ast}\Omega_{2}F_{4}}{G}\left(  e^{iK_{-}%
z}-e^{iK_{+}z}\right)
\end{array}
\text{(5)}%
\]

where $K_{\pm}=\frac{\omega}{c}+\frac{\kappa_{03}D_{p}+\kappa_{02}D_{m}}%
{2D}\pm \frac{\sqrt{G}}{2D}$, and
\[%
\begin{array}
[c]{l}%
D=F_{1}F_{2}F_{3}F_{4}+\left \vert \Omega_{2}\right \vert ^{2}\left \vert
\Omega_{c}\right \vert ^{2}-F_{3}F_{4}\left \vert \Omega_{2}\right \vert ^{2}\\
-F_{2}F_{4}\left \vert \Omega_{1}\right \vert ^{2}-F_{1}F_{2}\left \vert
\Omega_{c}\right \vert ^{2}\\
D_{m}=F_{4}\left \vert \Omega_{1}\right \vert ^{2}+F_{1}\left \vert \Omega
_{c}\right \vert ^{2}-F_{1}F_{3}F_{4}\\
D_{p}=F_{4}\left \vert \Omega_{2}\right \vert ^{2}-F_{1}F_{2}F_{4}\\
G=\left(  \kappa_{02}D_{m}-\kappa_{03}D_{p}\right)  ^{2}+4\kappa_{03}%
\kappa_{02}\left \vert \Omega_{1}\right \vert ^{2}\left \vert \Omega
_{2}\right \vert ^{2}F_{4}^{2}%
\end{array}
\]

with $F_{1}=\omega+\Delta_{1}+i\gamma_{1}$, $F_{2}=\omega+\Delta_{2}%
+i\gamma_{2}$, $F_{3}=\omega+\Delta_{p}+i\gamma_{3}$, $F_{4}=\omega+\Delta
_{c}+i\gamma_{4}$.

In what follows, as discussed in Refs. [15,17], the FWM conversion efficiency
can be defined as,%

\begin{equation}
\eta=\frac{\omega_{m}\kappa_{02}\kappa_{03}\left \vert \Omega_{1}\right \vert
^{2}\left \vert \Omega_{2}\right \vert ^{2}F_{4}^{2}}{\omega_{p}G}\exp \left \{
-2\operatorname{Im}\left[  K_{+}\left(  0\right)  \right]  L\right \}  .
\tag{6}%
\end{equation}

\section{Numerical results}

Now, if we choose the initial incident pulse $\Omega_{p}\left(  0,t\right)
=\Omega_{p}\left(  0,0\right)  \exp \left(  -t^{2}/\tau^{2}\right)  $, and we
can easily obtain $\Lambda_{p}\left(  0,\omega \right)  =\Omega_{p}\left(
0,0\right)  \tau \sqrt{\pi}\exp \left[  -\left(  \omega \tau \right)
^{2}/4\right]  $ ($\tau$ the is the pulse width). It is well known that the
$\left \vert \Lambda_{m}\left(  z,\omega \right)  \right \vert $ can be used to
calculate the generated FWM intensity. Some numerical results about the
relative intensity of the generated FWM field $\left \vert \Lambda_{m}\left(
z,\omega \right)  /\Omega_{p}\left(  0,0\right)  \tau \sqrt{\pi}\right \vert $ (a
dimensionless quantity) and the FWM conversion efficiency $\eta$\ will be
given in the Figures 2-4. In the following numerical calculations, we choose
the parameters to be dimensionless units by scaling $\gamma=\gamma_{3}$. The
choice of some parameters might be not very reasonably in our paper, however,
via properly adjusting other physical variables, we believe that some
experimental scientists have adequate wisdom to deal with this problem.

In the Figure 2, we plot the relative intensity of the generated FWM field
$\left \vert \Lambda_{m}\left(  z,\omega \right)  /\Omega_{p}\left(  0,0\right)
\tau \sqrt{\pi}\right \vert $ as a function of $\omega$\ for different EIT
windows. It can be easily seen from Fig. 2 that the relative intensities of
the generated FWM field change dramatically for different EIT windows. The
reason for the above result can be qualitatively explained as follows. Our
atomic system is a kind of atomic configuration which owns the property of the
double-dark resonances [21,22]. Under this condition, the coherent interaction
for this atomic system can lead to the emergence of sharp spectral features of
interacting double-dark resonances which strongly modify the optical
properties of the double EIT windows. Therefore, it can be seen that the FWM
generated waves are clearly different for different EIT windows at that time.

In the following numerical calculations, we shows the effect of the intensity
of the external coherent driving field on the FWM conversion efficiency $\eta
$\ in the Figure 3. We find that, when the intensity of the external coherent
driving field is small, the FWM conversion efficiency is low. However, with
the increasing the intensity of the external coherent driving field, the FWM
conversion efficiency becomes very large. The maximum FWM efficiency is
greater than 25\%, which is orders of magnitude larger than previous schemes
based on the two electromagnetically induced transparency. In order to test
the validity of the analysis described above, we carry out extensive numerical
calculations in the Figure 4. We give the three-dimensional plot of the
dependence of the FWM conversion efficiency $\eta$ on the\ two laser pump
fields $\Omega_{1}$ and $\Omega_{2}$. Clearly, the three-dimensional plot is
also verified the above comments.

Before ending this section, we would like to mention the two key points of the
present study. One of the major differences between our scheme with those
previous studies is that the maximum FWM efficiency (nearly 30\%) is larger
than previous schemes based on the two electromagnetically induced
transparency, and the high efficiency Four-Wave Mixing is induced by a easily
controlled coherent driving field. The second point is that the effects of the
intensity of the external coherent driving field on the FWM conversion
efficiency $\eta$ is monotonically when the coherent driving field is stronge
(see Figs. 3-4). A reasonable explanation for this is that, With the presence
of the external coherent driving field, we can obtain two quantum interference
channels. with the increasing the intensity of the external coherent driving
field, the destructive and constructive interferences induced via the two
quantum interference channels will keep a balance, so the intensity of the
coherent driving field will influence the FWM conversion efficiency $\eta$
monotonically at that time. In fact, this is a consequence of the competition
between the two quantum interference channels under relevant parametric conditions.

\section{Conclusions}

To sum up, We have analyzed a five-level $\wedge$-configuration Four-Wave
Mixing (FWM) scheme for obtaining a high-efficiency FWM based on the two
electromagnetically induced transparency. We find that the maximum FWM
efficiency is nearly 30\%, which is orders of magnitude larger than previous
schemes based on the two electromagnetically induced transparency. We hope
that our results may be helpful for experimental studies.

We would like to thank Prof. Hongyi Fan and Prof. Aixi Chen for helpful
discussion and encouragement.

\section{References}

\begin{itemize}
\item[{[1]}] Harris S E 1997 Phys. Today \textbf{50 }36

\item[{[2]}] Harris S E and Hau L V 1999 Phys. Rev. Lett. \textbf{82 }4611

\item[{[3]}] Lukin M D and Imamoglu A 2000 Phys. Rev. Lett. \textbf{84} 1419

\item[{[4]}] Wu Y, Wen L and Zhu Y 2003 Opt. Lett. \textbf{28} 631

\item[{[5]}] Zhu Y, Saldana J, Wen L and Wu Y 2004 J. Opt. Soc. Am. B
\textbf{21} 806

\item[{[6]}] Deng L and Payne M G 2003 Phys. Rev. Lett. \textbf{91} 243902

\item[{[7]}] Payne M G and Deng L 2003 Phys. Rev. Lett. \textbf{91} 123602

\item[{[8]}] Wu Y and Deng L 2004 Opt. Lett. \textbf{29} 1144

\item[{[9]}] Wang J, Zhu Y, Jiang K J and Zhan M S 2003 Phys. Rev. A
\textbf{68} 063810

\item[{[10]}] Zibrov A S, Ye C Y, Rostovsev Y V, Matsko A B and Scully M O
2002 Phys. Rev. A \textbf{65} 043817

\item[{[11]}] Matsko A B, Novikova I, Welch G R and Zubairy M S 2003 Opt.
Lett. \textbf{28} 96

\item[{[12]}] Harris S E, Field J E and Imamoglu A 1990 Phys. Rev. Lett.
\textbf{64} 1107

\item[{[13]}] Deng L, Kozuma M, Hagley E W and Payne M G 2002 Phys. Rev. Lett.
\textbf{88} 143902

\item[{[14]}] Wu Y, Saldana J and Zhu Y 2003 Phys. Rev. A \textbf{67} 013811

\item[{[15]}] Wu Y, Payne M G, Hagley E W and Deng L 2004 Opt. Lett.
\textbf{29} 2294

\item[{[16]}] Niu Y, Li R and Gong S 2005 Phys. Rev. A \textbf{71} 043819

\item[{[17]}] Li H J, Huang G X 2007 Phys. Rev. A \textbf{76} 043809

\item[{[18]}] Zhang Y, Brown A W and Xiao M 2007 Phys. Rev. Lett. \textbf{99} 123603

\item[{[19]}] Wu Y and Yang X 2005 Phys. Rev. A \textbf{71} 053806

\item[{[20]}] Wu Y and Yang X 2004 Phys. Rev. A \textbf{70} 053818

\item[{[21]}] Paspalakis E, Kylstra N J and Knight P L 2002 Phys. Rev. A
\textbf{65} 053808;

Paspalakis E and Knight P L 2002 Phys. Rev. A \textbf{66} 015802

\item[{[22]}] Goren C, Wilson-Gordon A D, Rosenbluh M and Friedmann H 2004
Phys. Rev. A \textbf{69} 063802
\end{itemize}

\bigskip

\bigskip

\begin{figure}[ptb]
\label{fig1} \centering \includegraphics[width=10cm]{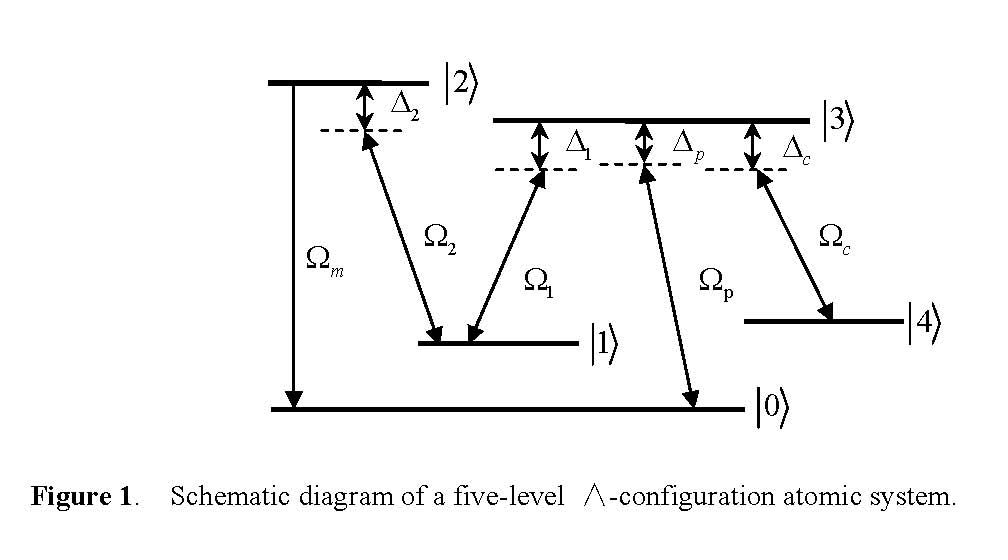}\caption{{}}%
\end{figure}

\begin{figure}[ptb]
\label{fig2} \centering \includegraphics[width=10cm]{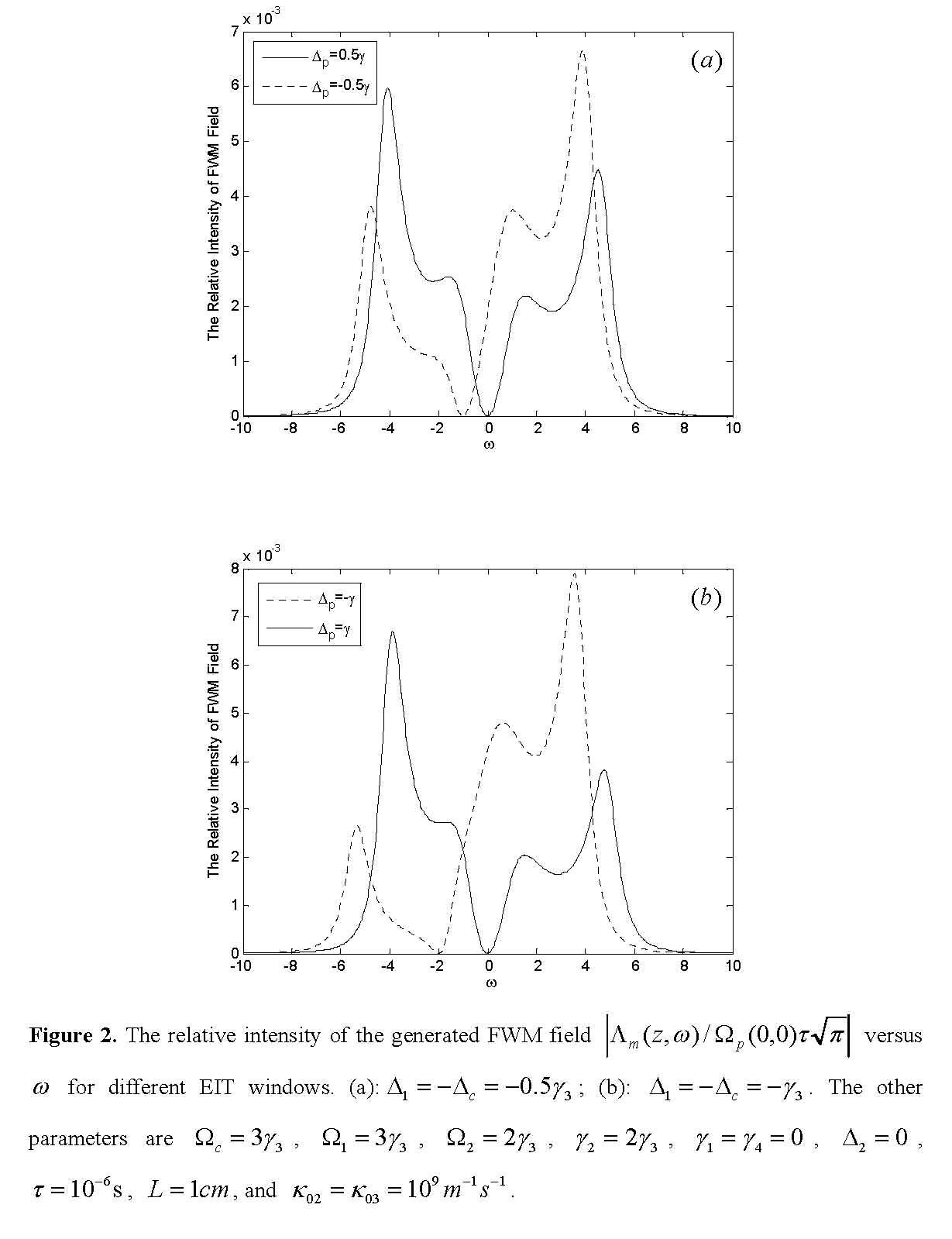}\caption{{}}%
\end{figure}\begin{figure}[ptb]
\label{fig3} \centering \includegraphics[width=10cm]{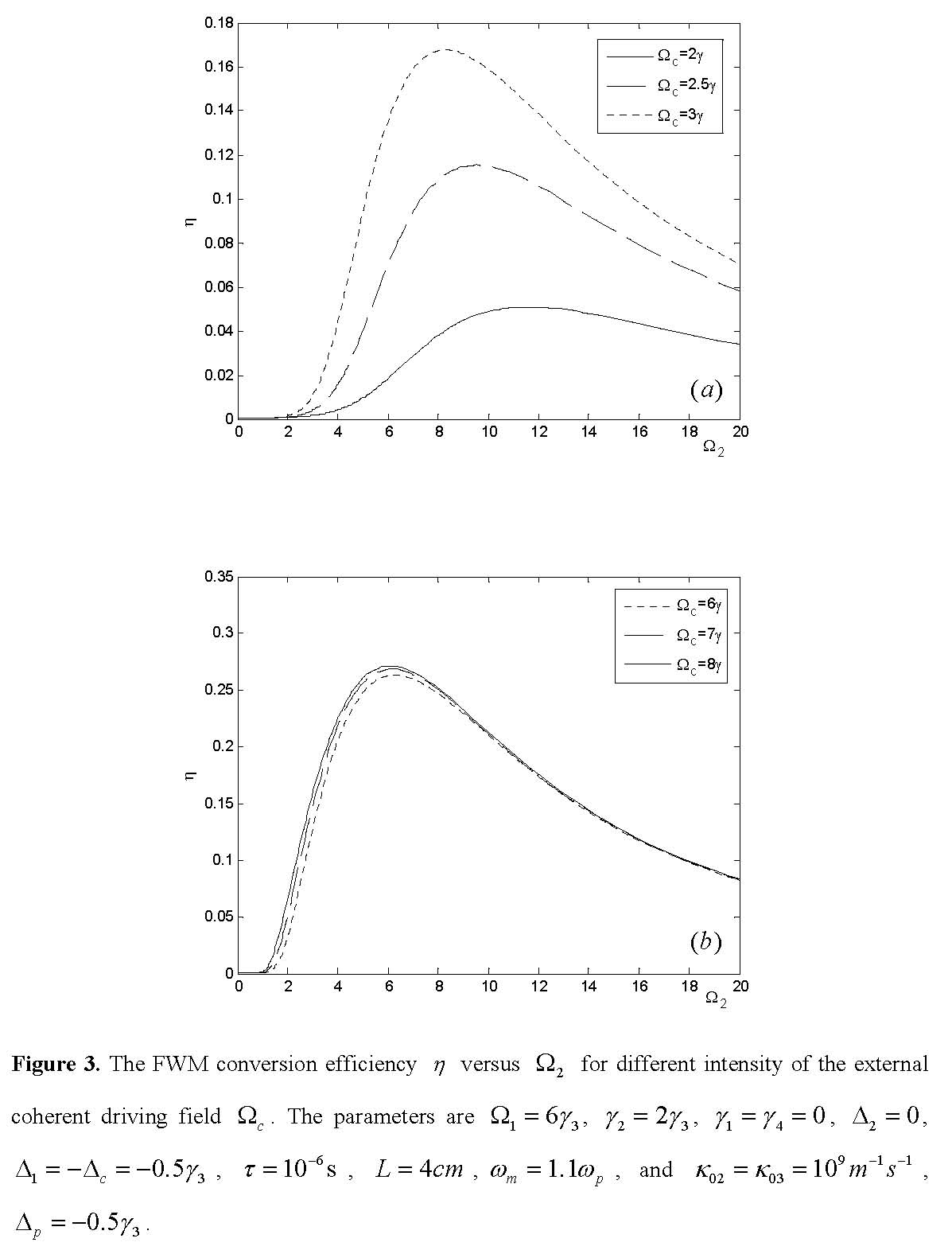}\caption{{}}%
\end{figure}\begin{figure}[ptb]
\label{fig4} \centering \includegraphics[width=10cm]{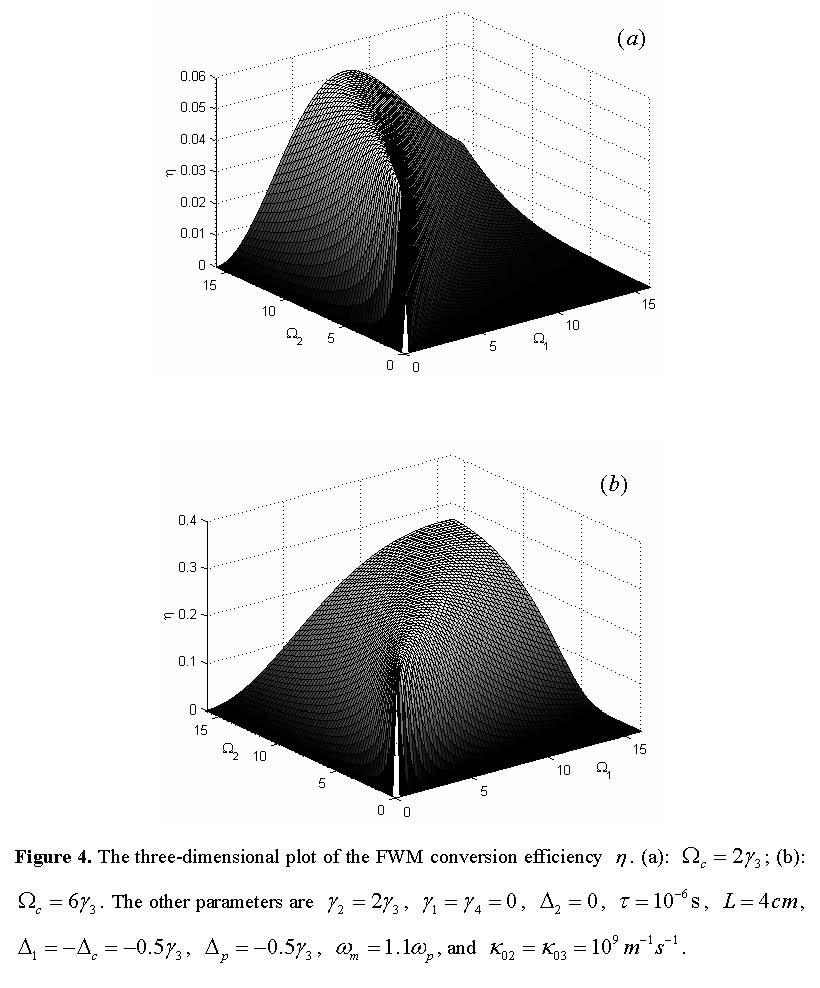}\caption{{}}%
\end{figure}

\end{document}